\documentclass[aps,prl,showpacs,amsmath,amssymb,sort&compress,comma,nofootinbib,floatfix,twocolumn,superscriptaddress]{revtex4-1}
\usepackage{graphicx}
\usepackage{color}
\usepackage{subfigure}
\usepackage[dvips,hyperfootnotes=false,breaklinks=true]{hyperref}
\hypersetup{
bookmarks=true,
colorlinks=true,
linkcolor=blue,
anchorcolor=blue,
citecolor=blue,
filecolor=blue,
urlcolor=blue,
bookmarksnumbered=true,
}

\begin{document}

\newcommand{\tbk}{{\tilde{\bf k}}\/}
\newcommand{\bk}{{\bf k}}
\newcommand{\bQ}{{\bf Q}}
\newcommand{\tbkp}{\tilde{\bf k}'}
\newcommand{\etal}{{\it et al.}\/}
\newcommand{\gtwid}{\mathrel{\raise.3ex\hbox{$>$\kern-.75em\lower1ex\hbox{$\sim$}}}}
\newcommand{\ltwid}{\mathrel{\raise.3ex\hbox{$<$\kern-.75em\lower1ex\hbox{$\sim$}}}}
\newcommand{\ut}[1]{\mathrm{\; #1}}

\title{Glide-Plane Symmetry and Superconducting Gap Structure of Iron-Based Superconductors}

\author{Y. Wang}
\affiliation{Center for Nanophase Materials Sciences, Oak Ridge National Laboratory, Oak Ridge, Tennessee 37831, USA}
\affiliation{Department of Physics and Astronomy, University of Tennessee, Knoxville, Tennessee 37996, USA}
\affiliation{Department of Physics, University of Florida, Gainesville, Florida 32611, USA}

\author{T. Berlijn}
\affiliation{Center for Nanophase Materials Sciences, Oak Ridge National Laboratory, Oak Ridge, Tennessee 37831, USA}
\affiliation{Computer Science and Mathematics Division, Oak Ridge National Laboratory, Oak Ridge, Tennessee 37831, USA}

\author{P. J. Hirschfeld}
\affiliation{Department of Physics, University of Florida, Gainesville, Florida 32611, USA}

\author{D. J. Scalapino}
\affiliation{Department of Physics, University of California, Santa Barbara, California 93106-9530, USA}

\author{T. A. Maier}
\affiliation{Center for Nanophase Materials Sciences, Oak Ridge National Laboratory, Oak Ridge, Tennessee 37831, USA}
\affiliation{Computer Science and Mathematics Division, Oak Ridge National Laboratory, Oak Ridge, Tennessee 37831, USA}

\begin{abstract}
We consider the effect of glide-plane symmetry of the Fe-pnictogen/chalcogen layer in Fe-based
superconductors on pairing in spin fluctuation models. Recent theories have proposed that so-called
$\eta$-pairing states with nonzero total momentum can be realized and possess exotic properties such as odd
parity spin singlet symmetry and time-reversal symmetry breaking. Here we show that $\eta$ pairing is
inevitable when there is orbital weight at the Fermi level from orbitals with even and odd mirror
reflection symmetry in $z$; however, by explicit calculation, we conclude that the gap function that
appears in observable quantities is identical to that found in earlier, 1 Fe per unit cell pseudocrystal
momentum calculations.
\end{abstract}


\maketitle

The common element in the crystal structure of all Fe-based superconductors is a two-dimensional plane of Fe
atoms on a square lattice with pnictogen/chalcogen atoms sitting in alternating positions below or above the
center of each square \cite{Stewart:2011kw,Wen:2011hn,Hirschfeld2011,Dai:2012em}. The alternating buckling of
pnictogen/chalcogen atoms results in a unit cell with two inequivalent Fe atoms. A model that takes into
account all Fe $d$ orbitals therefore has ten orbitals (five $d$-orbitals per Fe). In spite of this, most
theoretical calculations (e.g., Refs.~\onlinecite{Kuroki2008,Graser2009} and many others) have been carried
out using a five-orbital model for an ``unfolded'' Brillouin zone (BZ) of the 1 Fe unit cell, which appears to
miss the effects of the out-of-plane pnictogen/chalcogen degrees of freedom on the band structure. In
particular, an important question has been raised regarding whether these five-orbital calculations can
correctly determine the superconducting properties such as the gap structure, given the large
nonperturbative effects of the pnictogen/chalcogen potential that appear to be neglected in these studies.
In addition there are questions regarding the possibility of odd parity spin
singlet~\cite{Hu2012,Hu2013,Hao2014} and time-reversal breaking~\cite{Lin2014} associated with the so-called
$\eta$ pairing~\cite{Hu2012,Hu2013,Hao2014,Lin2014,KhodasChubukovPRL2012}.

To better understand this issue, we review the implications of the glide-plane symmetry of a single
Fe-pnictogen/chalcogen plane, as has been discussed before by various
authors~\cite{Lee2008,Eschrig2009,Andersen11,Casula_Sorella_2013,Fischer:2013hx,Cvetkovic2013}: While the 1
Fe lattice does not have translational symmetry since the two sublattices A and B made up, respectively, of
the two different Fe atoms are inequivalent, it is symmetric under the glide-plane symmetry operation
$P_z=T_r\sigma_z$, i.e., a one unit translation along the $x$- or $y$-direction $T_r$ combined with a
reflection $\sigma_z$ along $z$. As a consequence, the diagonal intrasublattice hopping between $d$ orbitals
that are even under $P_z$ ($xy,\,x^2-y^2,\,3z^2-r^2$) and orbitals that are odd ($xz,\,yz$) changes sign
between the A and B sublattice. When transformed to the physical 1 Fe crystal momentum $\bk$ space, this
leads to a mixing between momenta $\bk$ and $\bk+\bQ$ with $\bQ=(\pi,\pi)$ of the type $\sum_{\bk,\sigma}
\left[t^{xz,xy}(\bk)c^\dagger_{xz,\sigma, \bk+\bQ}c^{\phantom\dagger}_{xy,\sigma, \bk}+{\rm H.c.}\right]$,
and other similar terms between even and odd (with respect to $P_z$) orbitals. Because of this, there are
off-diagonal propagators involving even and odd orbitals with momenta $\bk$ and $\bk+\bQ$. This has important
consequences with respect to the pairing: in addition to the standard zero center of mass momentum pairs
$\langle c_{\ell_1,\uparrow,\bk} c_{\ell_2,\downarrow,-\bk}\rangle$ for $\ell_1,\ell_2$ either both even or
both odd orbitals, there are also nonzero total momentum $\eta$ pairs $\langle c_{\ell_1,\uparrow,\bk}
c_{\ell_2,\downarrow,-\bk+\bQ}\rangle$ for $\ell_1$ even, $\ell_2$ odd or vice versa~\cite{Lin2014}.

\begin{figure}
	\includegraphics[width=8cm]{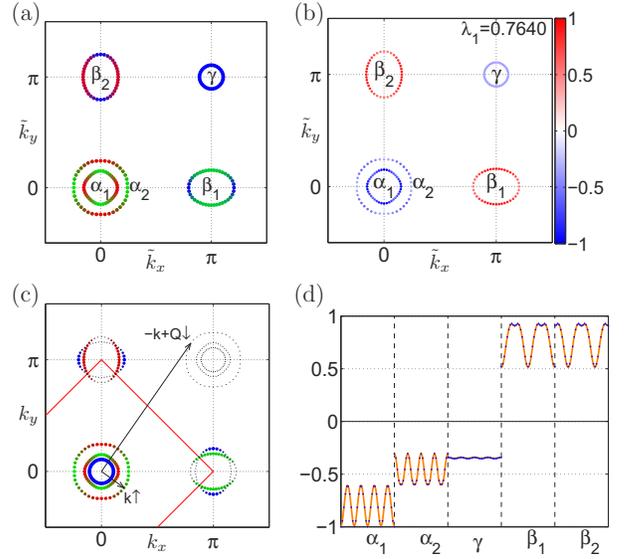}

\caption{(color online). (a) Fermi surfaces and (b) the leading gap function for the five-orbital model in
the zone of the pseudocrystal momentum $\tbk$. The Fermi surface is colored to show the dominant orbital
weight [$d_{xz}$ red (gray), $d_{yz}$ green (light gray), $d_{xy}$ blue (dark gray)]. (c) The unfolded Fermi
surface in the physical crystal momentum $\bk$-space. The size of the dots is proportional to the sum of the
orbital weights of the spectral function and the color shows the dominant orbital weight. The red line
denotes the boundary of the 2 Fe per unit cell Brillouin zone. An ``$\eta$'' pair ($\bk\uparrow$,
$-\bk+\bQ\downarrow$) is shown. (d) Comparison of the angle dependence of the leading gap function calculated
from the five-orbital model (orange line) and the ten-orbital model (blue dots) on the various Fermi pockets.
Here we denote the two crossed electron pockets in the ten-orbital model at the $X$ point as $\beta_1$ and
$\beta'_1$ and at the $Y$ point $\beta_2$ and $\beta'_2$. The gaps along the $\beta'_{1,2}$ pockets are not
plotted since $\beta'_1=\beta_2$ and $\beta'_2=\beta_1$ by symmetry.}
  \label{fig1}
\end{figure}

However, this mixing is absent if one uses the eigenvalues of $P_z$, i.e., the pseudocrystal momentum $\tbk$,
to classify the states~\cite{Lee2008,Eschrig2009,Andersen11}. This basically corresponds to shifting the
momentum of either the even or the odd orbitals by $\bQ$. Here we choose the shift in the even orbitals so
that states defined in the pseudocrystal momentum $\tbk$ are related to the states defined with the physical
crystal momentum $\bk$ through
\begin{align}
	\label{eq:pcm} \tilde{c}_{\ell,\sigma,\tbk} =
	\begin{cases}
		c_{\ell,\sigma,\bk}, & \text{if $\ell$ odd,} \\
		c_{\ell,\sigma,\bk+\bQ}, & \text{if $\ell$ even.}
	\end{cases}
\end{align}
The Hamiltonian is diagonal in $\tbk$ and, as we will discuss, the usual five-orbital calculations, when
performed in this space, automatically take into account the additional terms stemming from the mixing
between $\bk$ and $\bk+\bQ$ in the physical 1 Fe crystal momentum $\bk$ space. Here $\eta$ pairing is
implicitly included since pairs like $\langle {\tilde c}_{xy,\uparrow,\tbk} {\tilde c}_{xz,\downarrow,-\tbk}
\rangle$ in $\tbk$ space transform to $\langle c_{xy,\uparrow,\bk} c_{xz,\downarrow,-\bk+\bQ} \rangle$ in
$\bk$ space as indicated in Fig.~\ref{fig1}(c). Here we study the parity properties of these terms, the way
in which they combine with normal (zero center of mass momentum) pairing states, and their implications for
the gap structure in the physical crystal momentum $\bk$ space. We calculate the one-particle spectral
function in the proper crystal momentum space and show that the energy gaps deduced from spectral function
leading edges correspond to those calculated in the 1 Fe zone, although the quasiparticle weights are
strongly renormalized. We conclude that, as usual, an even frequency gap for a singlet pair has even parity
in the band basis and there is no time-reversal symmetry breaking as a result of $\eta$ pairing.

To this end, we use the 2D five-orbital tight-binding model for LaOFeAs introduced in Graser \textit{et
al}.~\cite{Graser2009}. This model was obtained from a Wannier transformation of a local density
approximation band structure calculation of this compound with a 2 Fe ten-orbital model and performing a
gauge transformation corresponding to a $\pi$-phase shift of the even orbitals on the B sublattice, which in
momentum space corresponds to a transformation to pseudocrystal $\tbk$ momentum. The Fermi surface of this
model in the 1 Fe pseudocrystal momentum $\tbk$ space is shown in Fig.~\ref{fig1}(a), with the dominant
orbital weights indicated by the coloring. The corresponding Fermi surface in physical crystal momentum
$\bk$-space is plotted in Fig.~\ref{fig1}(c). According to Eq.~(\ref{eq:pcm}), it is obtained by shifting the
even orbital contribution by $\bQ$. The size of the points indicates the sum of the orbital weights.

This model is then supplemented with the usual Hubbard (intraorbital $U$ and interorbital $U'$) and Hund
(Hund's rule coupling $J$ and pair hopping $J'$) interactions. Here we assume spin rotational invariance so
that $U'=U-2J$ and $J'=J$; set $U=1.3\,\text{eV}$ and $J=0.2\,\text{eV}$, and take $\langle n\rangle =5.95$.
We then use a random-phase approximation to calculate the pairing interaction
$\Gamma_{\ell_1\ell_2\ell_3\ell_4}(\tbk,\tbk')$ which represents the particle-particle scattering of
electrons in orbitals $\ell_1,\ell_4$ with momenta $(\tbk,-\tbk)$ to electrons in orbitals $\ell_2,\ell_3$
with momenta $(\tbk',-\tbk')$. The pairing strengths $\lambda_\alpha$ for various pairing channels $\alpha$
are then given as the eigenvalues of
\begin{align}
  -\sum_j\oint_{C_j}\frac{d\tbk'_\parallel}{(2\pi)^2 v_F(\tbk'_\parallel)}
    \Gamma_{ij}(\tbk,\tbk') g_\alpha(\tbk') = \lambda_\alpha g_\alpha(\tbk)\,.
\end{align}
Here, $\Gamma_{ij}(\tbk,\tbk')$ represents the irreducible vertex for the scattering of a pair of electrons
$(\tbk\uparrow,-\tbk\downarrow)$ on Fermi pocket $C_i$ to $(\tbk'\uparrow,-\tbk'\downarrow)$ on pocket $C_j$.
It is obtained from $\Gamma_{\ell_1\ell_2\ell_3\ell_4}(\tbk,\tbk')$ as
\begin{eqnarray}
  \label{eq:Gammac}
  \Gamma_{ij}(\tbk,\tbk') = &&\sum_{\ell_1,\ell_2,\ell_3,\ell_4}
    {\tilde a}^{\ell_1^*}_{\nu,\tbk}{\tilde a}^{\ell_4^*}_{\nu,-\tbk}
    \Gamma_{\ell_1\ell_2\ell_3\ell_4}(\tbk,\tbk')\nonumber\\
	&& \hspace{1.5cm}\times
    {\tilde a}^{\ell_2}_{\mu,\tbk'}{\tilde a}^{\ell_3}_{\mu,-\tbk'}\,,
\end{eqnarray}
where the matrix elements ${\tilde a}^{\ell}_{\nu,\tbk}=\langle
\widetilde{\ell\bk}|\widetilde{\nu\bk}\rangle$ transform the orbital basis to the band representation in
pseudocrystal momentum space. The momenta $\tbk$ and $\tbk'$ in Eq.~(\ref{eq:Gammac}) are restricted to the
Fermi surface and $v_F(\tbk'_\parallel)$ is the Fermi velocity. The eigenfunction $g_\alpha(\tbk)$ for the
largest eigenvalue determines the leading pairing instability and provides an approximate form for the
superconducting gap $\tilde{\Delta}(\tbk) \propto g_\alpha(\tbk)$. The structure of the leading gap function
$g_\alpha(\tbk)$ with $s_\pm$-wave symmetry on the Fermi surface is shown in Fig.~\ref{fig1}(b).

We have also calculated the leading gap function and the eigenvalue in the original ten-orbital model, from
which the five-orbital model was derived through a gauge transformation, as discussed above. We obtain the
same leading eigenvalue $\lambda=0.76$ in the ten-orbital as in the five-orbital model, and
Fig.~\ref{fig1}(d) shows that the gap function obtained in the ten-orbital model is identical to what is
obtained in the five-orbital model. From this it is clear that calculations performed in the 1 Fe
five-orbital pseudocrystal momentum space indeed contain all information of the more complex ten-orbital
calculation performed in the 2 Fe crystal momentum space.

As discussed above, this includes the information about $\eta$-pairing terms in the physical 1 Fe momentum
$\bk$ space. In order to analyze the structure of these terms, we transform the gap function
$\tilde{\Delta}_\nu(\tbk)$ that we obtained in pseudocrystal momentum space to 1 Fe physical crystal momentum
$\bk$ space. To this end, we first transform $\tilde{\Delta}_\nu(\tbk)$ from band to orbital space and then
to $\bk$ space. This gives normal pairing terms with zero center of mass momentum,
\begin{align}
  \label{eq:Delta_N_orb}
  \langle c_{\ell_1\uparrow,\bk}&c_{\ell_2\downarrow,-\bk}-c_{\ell_1\downarrow,\bk}c_{\ell_2\uparrow,-\bk}\rangle
  \propto \Delta^N_{\ell_1\ell_2}(\bk) =\\
	& \begin{cases}
		\tilde{a}^{\ell_1}_{\nu,\bk}\tilde{a}^{\ell_2}_{\nu,-\bk}\tilde{\Delta}_\nu(\bk),
                    & \text{$\ell_1,\ell_2$ odd,}\\
		\tilde{a}^{\ell_1}_{\nu,\bk-{\bf Q}}\tilde{a}^{\ell_2}_{\nu,-\bk+{\bf Q}}\tilde{\Delta}_\nu(\bk-{\bf Q}),
                    & \text{$\ell_1,\ell_2$ even,}\\
		0,          & \text{otherwise,}\nonumber
	\end{cases}
\end{align}
and $\eta$-pairing terms with center of mass momentum $\bQ$,
\begin{align}
  \label{eq:Delta_eta_orb}
  \langle c_{\ell_1\uparrow,\bk}&c_{\ell_2\downarrow,-\bk+{\bf Q}}-c_{\ell_1\downarrow,\bk}c_{\ell_2\uparrow,-\bk+{\bf Q}}\rangle
  \propto \Delta^\eta_{\ell_1\ell_2}(\bk) =\\
	& \begin{cases}
		\tilde{a}^{\ell_1}_{\nu,\bk}\tilde{a}^{\ell_2}_{\nu,-\bk}\tilde{\Delta}_\nu(\bk),
                    & \text{$\ell_1$ odd, $\ell_2$ even,}\\
		\tilde{a}^{\ell_1}_{\nu,\bk-{\bf Q}}\tilde{a}^{\ell_2}_{\nu,-\bk+{\bf Q}}\tilde{\Delta}_\nu(\bk-{\bf Q}),
                    & \text{$\ell_1$ even, $\ell_2$ odd,}\\
		0,          & \text{otherwise.}\nonumber
	\end{cases}
\end{align}
Here and in the following we have replaced $\tbk$ by $\bk$ for odd and $\bk+\bQ$ for even orbitals $\ell$, as
noted in Eq.~(\ref{eq:pcm}). For a given Fermi momentum $\bk$, $\nu$ labels the band that crosses the Fermi
energy at $\bk$.
\begin{figure}
	\includegraphics[width=8cm]{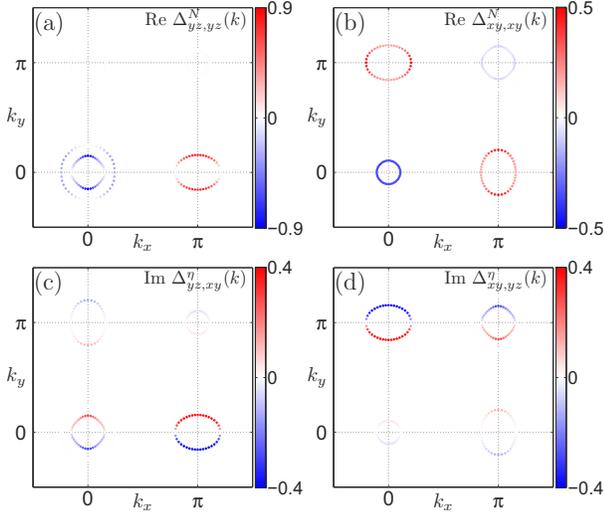}

\caption{(color online). Gap functions $\Delta_{\ell_1\ell_2}(\bk)$ in the orbital basis plotted on the Fermi
surface in the physical momentum BZ. (a)--(b) When $\ell_1$ and $\ell_2$ have the same $z$-reflection
symmetry, one has a normal ($\bk$,$-\bk$) pairing and the gap function $\Delta^{N}_{\ell_1\ell_2}(\bk)$ is
real and has even parity. (c)--(d) When $\ell_1$ and $\ell_2$ have different $z$-reflection symmetry, one has
a ($\bk$,$-\bk+\bQ$) $\eta$ pairing and $\Delta^{\eta}_{\ell_1\ell_2}(\bk)$ is purely imaginary and has odd
parity.}
  \label{fig2}
\end{figure}

Figure~\ref{fig2} shows the gap functions $\Delta_{\ell_1\ell_2}(\bk)$ for four different combinations of
orbitals $\ell_1$ and $\ell_2$. In Figs.~\ref{fig2}(a) and \ref{fig2}(b), one sees that when $\ell_1$ and
$\ell_2$ have the same $z$-reflection symmetry, one has a normal ($\bk$,$-\bk$) pairing and the gap function
$\Delta^{N}_{\ell_1\ell_2}(\bk)$ is real and has even parity, i.e.,
$\Delta^{N}_{\ell_1\ell_2}(-\bk)=\Delta^{N}_{\ell_1\ell_2}(\bk)$. In contrast, when $\ell_1$ and $\ell_2$
have different $z$-reflection symmetry, one has a ($\bk$,$-\bk+\bQ$) $\eta$ pairing. In this case,
$\Delta^{\eta}_{\ell_1\ell_2}(\bk)$ is purely imaginary and has odd parity, i.e.,
$\Delta^{\eta}_{\ell_1\ell_2}(-\bk)=-\Delta^{\eta}_{\ell_1\ell_2}(\bk)$. Note that these gaps have the same
behavior under mirror reflections as the orbitally resolved gaps of Casula and Sorella
\cite{Casula_Sorella_2013}. Here we stress that the odd parity and the imaginary nature of these terms in
orbital space arises entirely from the product of matrix elements
$\tilde{a}^{\ell_1}_{\nu,\bk}\tilde{a}^{\ell_2}_{\nu,-\bk}$ and, therefore, is merely a reflection of the
glide-plane symmetry of the Fe-pnictogen/chalcogen plane. It does not reflect any exotic behavior of the
pairing interaction. We also point out that time-reversal symmetry requires that $\Delta_{\ell_1\ell_2}({\bf
k})=\Delta_{\ell_1\ell_2}^*(-{\bf k})$ for both the normal and $\eta$ gaps in Eqs.~({\ref{eq:Delta_N_orb})
and (\ref{eq:Delta_eta_orb}), respectively. Because the normal gap $\Delta^N_{\ell_1\ell_2}(\bk)$ has even
parity and is purely real, it satisfies time-reversal symmetry, as does the odd parity, purely imaginary
$\eta$-pairing gap $\Delta^\eta_{\ell_1\ell_2}(\bk)$. Both normal and $\eta$-pairing terms, however, coexist
in orbital space and contribute to the pairing condensate.

This raises the question of how these two terms combine, given their opposite parity. To study this, we
transform the gap back to band representation in physical crystal momentum $\bk$ space and obtain, for the
normal pairing,
\begin{align}
	\label{eq:Delta_N} \Delta^N_\nu(\bk) = \Delta^N_\text{odd}(\bk)+\Delta^N_\text{even}(\bk)\,,
\end{align}
where
\begin{subequations}
\begin{align}
  \Delta^N_\text{odd}(\bk) &= \sum_{\ell_1,\ell_2 \,\text{odd}}
    {\tilde a}_{\nu,\bk}^{\ell_1^*}{\tilde a}_{\nu,-\bk}^{\ell_2^*} \Delta^N_{\ell_1\ell_2}(\bk)\,,\\
  \Delta^N_\text{even}(\bk) &= \sum_{\ell_1,\ell_2 \,\text{even}}
    {\tilde a}_{\nu,\bk-\bQ}^{\ell_1^*}{\tilde a}_{\nu,-\bk+\bQ}^{\ell_2^*}\Delta^N_{\ell_1\ell_2}(\bk)\,.
\end{align}
\end{subequations}
Similarly, we obtain for the $\eta$-pairing terms,
\begin{align}
  \label{eq:Delta_eta}
  \Delta^\eta_\nu(\bk) = \Delta^\eta_\text{odd-even}(\bk)+\Delta^\eta_\text{even-odd}(\bk)\,,
\end{align}
where
\begin{subequations}
\label{eq:Delta_eta}
\begin{align}
 \Delta^\eta_\text{odd-even}(\bk) &= \mspace{-10mu}\sum_{\text{$\ell_1$ odd},\text{$\ell_2$ even}}\mspace{-10mu}
   {\tilde a}_{\nu,\bk}^{\ell_1^*}{\tilde a}_{\nu,-\bk}^{\ell_2^*} \Delta^\eta_{\ell_1\ell_2}(\bk)\,,\\
 \Delta^\eta_\text{even-odd}(\bk) &= \mspace{-10mu}\sum_{\text{$\ell_1$ even},\text{$\ell_2$ odd}}\mspace{-10mu}
   {\tilde a}_{\nu,\bk-\bQ}^{\ell_1^*}{\tilde a}_{\nu,-\bk+\bQ}^{\ell_2^*} \Delta^\eta_{\ell_1\ell_2}(\bk)\,.
\end{align}
\end{subequations}
Here we have used the fact that the matrix elements $a_{\nu,\bk}^{\ell}$, which provide the transformation
from the orbital to the band representation in physical crystal momentum $\bk$ space, are given by the matrix
elements in pseudocrystal momentum space, ${\tilde a}_{\nu,\bk}^{\ell}$ for $\ell$ denoting an odd orbital
and ${\tilde a}_{\nu,\bk-\bQ}^{\ell}$ for $\ell$ denoting an even orbital.
\begin{figure}
	\includegraphics[width=5cm]{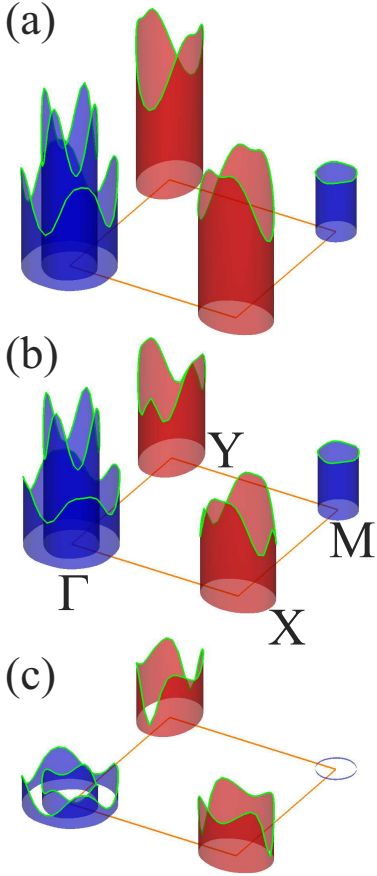}

\caption{(color online). (a) The leading gap function $\tilde{\Delta}(\tilde{\bk})$ in the band
representation calculated in the five-orbital model in pseudocrystal momentum space [red $=$ positive (gap
along pockets at $X$ and $Y$ points), blue $=$ negative (gap along pockets at $\Gamma$ and $M$ points)]. When
transformed to physical crystal momentum $\mathbf{k}$ space, the gap splits into the normal even-even and
odd-odd contributions $\Delta^N_\text{odd}(\bk)+\Delta^N_\text{even}(\bk+\bQ)$ plotted in (b) and the
even-odd and odd-even $\eta$ contributions
$\Delta^\eta_\text{odd-even}(\bk)+\Delta^\eta_\text{even-odd}(\bk+\bQ)$ shown in (c). In the band
representation, all contributions have even parity.}
  \label{fig3}
\end{figure}

Then, using Eqs.~(\ref{eq:Delta_N_orb}) and (\ref{eq:Delta_eta_orb}) one can show that the gap function
$\tilde{\Delta}_\nu(\bk)$ calculated in the five-orbital model in the pseudocrystal momentum representation
splits into normal and $\eta$-pairing terms in the physical crystal momentum space, i.e.,
\begin{align}
  \tilde{\Delta}_\nu(\bk)&=\Delta^N_{\rm odd}(\bk)+\Delta^N_{\rm even}(\bk+\bQ)\nonumber\\
  &+\Delta^\eta_\text{odd-even}(\bk)+\Delta^\eta_\text{even-odd}(\bk+\bQ)\,.
  \label{eq:Deltasum}
\end{align}
Note that the even terms have their momentum shifted by $\bQ$, so they appear on the same Fermi pockets as
the odd terms. Figure~\ref{fig3} shows a graphical representation of this relation by plotting a 3D
representation of $\tilde{\Delta}_\nu(\bk)$ in the top panel, its normal contribution
$\Delta^N_\text{odd}(\bk)+\Delta^N_\text{even}(\bk+\bQ)$ in the middle panel, and its $\eta$ contribution
$\Delta^\eta_\text{odd-even}(\bk)+\Delta^\eta_\text{even-odd}(\bk+\bQ)$ in the bottom panel. One sees that
after the transformation to band representation, the $\eta$-pairing term has even parity (and is real), just
like the normal pairing contribution. This results from the combination of the product of matrix elements
${\tilde a}_{\nu,\bk}^{\ell_1^*}{\tilde a}_{\nu,-\bk}^{\ell_2^*}$, which has odd parity and is purely
imaginary, with the odd parity and imaginary $\Delta^\eta_{\ell_1 \ell_2}(\bk)$ for odd-even combinations of
$\ell_1$ and $\ell_2$. Thus, as usual, an even frequency gap in the band basis has even parity for a singlet
pair. From Eqs.~(\ref{eq:Delta_N_orb})--(\ref{eq:Delta_eta}) it is straightforward to show that the normal
and $\eta$ contributions to the gap have the same sign and share the same nodal structure in the case of a
nodal gap~\cite{footnote}. Thus, while the $\eta$ contribution is significant in the amplitude of the total
gap, it does not affect its sign and nodal structure and therefore does not qualitatively alter the low
temperature thermodynamic properties. This is discussed in more detail in the Supplemental
Material~\cite{SuppMaterial}.

Finally, we calculate the spectral function
\begin{align}
  A(\bk,\omega)= \sum_{\ell,\nu}|\langle \ell\bk|\widetilde{\nu\bk} \rangle|^2 \tilde{A}_\nu(\tbk, \omega)\nonumber
\end{align}
as measured in angle-resolved photoemission spectroscopy (ARPES) experiments in the proper 1 Fe crystal
momentum $\bk$ space. Here,
$\tilde{A}_\nu(\tbk,\omega)=u_{\nu}^2({\tbk})\delta(\omega-E_\nu({\tbk}))+v_\nu^2({\tbk})\delta(\omega+E_\nu({\tbk}))$
is the BCS spectral function in the pseudocrystal momentum space with
$E_\nu({\tbk})=\sqrt{\epsilon^2_\nu(\tbk)+\tilde{\Delta}^2_\nu(\tbk)}$ and the BCS coherence factors
$u^2_{\nu}({\tbk})=[1+\epsilon_\nu(\tbk)/E_\nu({\tbk})]/2$ and $v^2_\nu({\tbk})=1-u^2_\nu({\tbk})$. Realizing
that
\begin{align}
	\langle \ell\bk|\widetilde{\nu\bk} \rangle =
	\begin{cases}
		\tilde{a}_{\nu,\bk}^{\ell} \delta_{\bk,\tbk}, &\text{$\ell$ odd,}\\
		\tilde{a}_{\nu,\bk-\bQ}^{\ell} \delta_{\bk-\bQ,\tbk}, &\text{$\ell$ even,}
	\end{cases}
\end{align}
one arrives at
\begin{align}
  \label{eq:Akw}
  A(\bk,\omega) = \sum_\nu
    &\left[ \sum_{\ell\,\text{odd}} |\tilde{a}^\ell_{\nu,\bk}|^2 \tilde{A}_\nu(\bk,\omega)\right.\nonumber\\
	&+\left.\sum_{\ell\,\text{even}}|\tilde{a}^\ell_{\nu,\bk-\bQ}|^2 \tilde{A}_\nu(\bk-\bQ,\omega)\right].
\end{align}
Thus, the superconducting gap that enters $A(\bk,\omega)$ as measured in ARPES experiments is given by the
gap function $\tilde{\Delta}_\nu(\tbk)$ calculated in the five-orbital 1 Fe zone in pseudocrystal momentum
space and no further transformation is necessary. $\tilde{\Delta}_\nu(\tbk)$ implicitly encodes the strong
symmetry breaking potential associated with the pnictogen/chalcogen atom. The gap $\tilde{\Delta}_\nu(\tbk)$
entering the first $\ell=\text{``odd''}$ term in Eq.~(\ref{eq:Akw}) is shown in Fig.~\ref{fig3}(a) while the
gap entering the second $\ell=\text{``even''}$ contribution which appears on the ``shadow'' pockets is
obtained by shifting the gap by $\bQ$. As in the normal
state~\cite{Ku10,Lv:2011ct,Lin:2011jv,Brouet2012,Kong14}, the spectral weight in the superconducting state
associated with each contribution is modulated by the orbital weights $|\tilde{a}^\ell_{\nu,\bk}|^2$ and
$|\tilde{a}^\ell_{\nu,\bk-\bQ}|^2$, respectively, and this weight can differ substantially between the main
and shadow pockets, as seen in Fig.~\ref{fig1}(c). The spectral functions for both the normal and the
superconducting states are discussed in more detail and shown in the Supplemental
Material~\cite{SuppMaterial}.

To summarize, we have carried out microscopic calculations of the superconducting gap structure in 1 Fe and 2
Fe per unit cell models and shown that $\eta$ pairing is an important ingredient in the superconducting
condensate. We have demonstrated that it contributes with the usual even parity symmetry in band space and
that time-reversal symmetry is preserved, in contrast to recent proposals in the literature. Finally we have
shown that the gap function, which appears in observable quantities, is identical to that found in earlier,
1~Fe per unit cell pseudocrystal momentum calculations.

The authors acknowledge their useful discussions with A. Chubukov, M. Khodas, and W. Ku. P.J.H. and Y.W. were
supported by Grant No. DOE DE-FG02-05ER46236, and T.B. was supported as a Wigner Fellow at the Oak Ridge
National Laboratory. A portion of this research was conducted at the Center for Nanophase Materials Sciences,
which is sponsored at Oak Ridge National Laboratory by the Scientific User Facilities Division, Office of
Basic Energy Sciences, U.S. Department of Energy. This research was supported in part by Kavli Institute for
Theoretical Physics under National Science Foundation Grant No. PHY11-25915.

\setcounter{figure}{0}

\makeatletter
\renewcommand{\thefigure}{S\@arabic\c@figure}
\makeatother

\begin{widetext}
\section{[Supplemental Information]}

\section*{S1. Single particle spectral function}

As discussed in the main text and according to Eq.~(\ref{eq:Akw}), the spectral function
$\tilde{A}(\bk,\omega)$ in pseudocrystal momentum space splits into two contributions when transformed to
physical crystal momentum space $\bk$, i.e.,
\begin{align}
	\tilde{A}(\bk,\omega) = A_\text{odd}(\bk,\omega) + A_\text{even}(\bk+\bQ,\omega)\,,
\end{align}
with
\begin{align}
	A_\text{odd}(\bk,\omega) &=
        \sum_\nu \sum_{\ell \,\text{odd}} |\tilde{a}_{\nu,\bk}^\ell|^2 \tilde{A}_\nu(\bk,\omega)\,,\\
	A_\text{even}(\bk+\bQ,\omega) &=
        \sum_\nu\sum_{\ell \,\text{even}} |\tilde{a}_{\nu,\bk}^\ell|^2 \tilde{A}_\nu(\bk,\omega)\,.
\end{align}
The first contribution has purely odd orbital character and appears at the same momentum $\bk$, while the
second contribution has even orbital character and is shifted by $\bQ=(\pi,\pi)$. Each contribution is
weighted by orbital matrix elements. This is illustrated in Fig.~\ref{fig:orbWt} where we plot the normal
state spectral function with $\omega=0$ on the Fermi surface. The odd-orbital contribution is depicted in
panel (a), while the $\bQ$ shifted even-orbital contribution is shown in (b).

\begin{figure}[!h]
  \centering
  \includegraphics[keepaspectratio=true,scale=1.0]{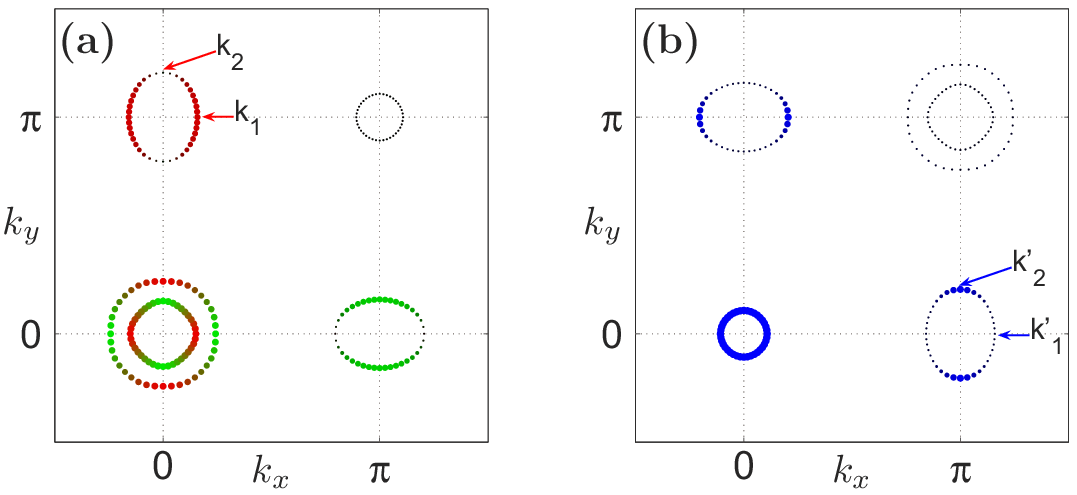}

\caption{The spectral function $A_\text{odd}(\bk,\omega=0)$ (a) and $A_\text{even}(\bk,\omega=0)$ (b) on the
Fermi surface in physical crystal momentum space where the color shows the dominant orbital weight ($d_{xz}$
red, $d_{yz}$ green and $d_{xy}$ blue). The size of the dots is proportional to $\sum_{\ell\,\rm{odd}}
|\tilde{a}^\ell_{\nu,\bk}|^2$ in (a) and $\sum_{\ell\,\rm{even}}|\tilde{a}^\ell_{\nu,\bk-\bQ}|^2$ in (b).}
  \label{fig:orbWt}
\end{figure}

The frequency dependence of the spectral function near $\omega=0$ is plotted in
Fig.~\ref{fig:LaOFeAs_n5.95_spectraWt} for momenta $\bk_1$ and $\bk_2$ on the ``odd'' parts of the Fermi
surface, and for momenta $\bk'_1$ and $\bk'_2$ on the ``even'' parts. The four $\bk$ points, $\bk_{1,2}$ and
$\bk'_{1,2}$, are indicated in Fig.~\ref{fig:orbWt}. These momenta are chosen so that $\bk'_1 = \bk_1+\bQ$
and $\bk'_2 = \bk_2+\bQ$ where $\bQ=(\pi,-\pi)$. The green triangles depict the spectral function in
pseudocrystal momentum space, $\tilde{A}(\bk_{1,2},\omega)$, while the red and blue solid lines show the two
contributions $A_\text{odd}(\bk_{1,2},\omega)$ and $A_\text{even}(\bk'_{1,2},\omega)$, respectively, in
physical crystal momentum space. The spectral functions in the normal state are plotted in panels (a) and
(d), while panels (b) and (e) are for the superconducting state, for which we have used the gap momentum
structure shown in Fig.~1(d) with a maximum gap amplitude of $5\ut{meV}$. The $\delta$-functions entering the
spectral functions are approximated by a Lorentzian, i.e., $\delta(\omega)\approx
\zeta/[\pi(\omega^2+\zeta^2)]$ and we have used an artificial broadening $\zeta=0.3\ut{meV}$. From these
plots one sees again how the spectral weights in the physical crystal momentum space at $\bk_1$ and $\bk'_1 =
\bk_1+\bQ$ add up to give the spectral weight in the pseudocrystal momentum space at $\bk_1$. In the
superconducting state, one sees that the leading edge gap in the ``odd'' part of the spectrum at momenta
$\bk_{1,2}$ is given by the gap $\tilde{\Delta}(\bk_{1,2})$ calculated in pseudocrystal momentum space, while
the gap in the ``even'' part of the spectrum at momenta $\bk'_{1,2}$ is given by
$\tilde{\Delta}(\bk'_{1,2}-\bQ)=\tilde{\Delta}(\bk_{1,2})$. To show the effect of the $\eta$ gap on the
spectral function, in Fig.~\ref{fig:LaOFeAs_n5.95_spectraWt} (c) and (f) we plot the spectral function at
$\bk_{1}$ ($\bk'_1$) and $\bk_{2}$ ($\bk'_2$), respectively, for the superconducting state with the $\eta$
contribution removed from the calculated gap function. As discussed in the main text, the normal and $\eta$
contributions to the gap have the same sign as $\tilde{\Delta}({\bf k})$. Because of the this, the gap in the
spectral function would be reduced if the $\eta$ contribution were removed.

\begin{figure}[!h]
  \centering
  \includegraphics[keepaspectratio=true,scale=1.0]{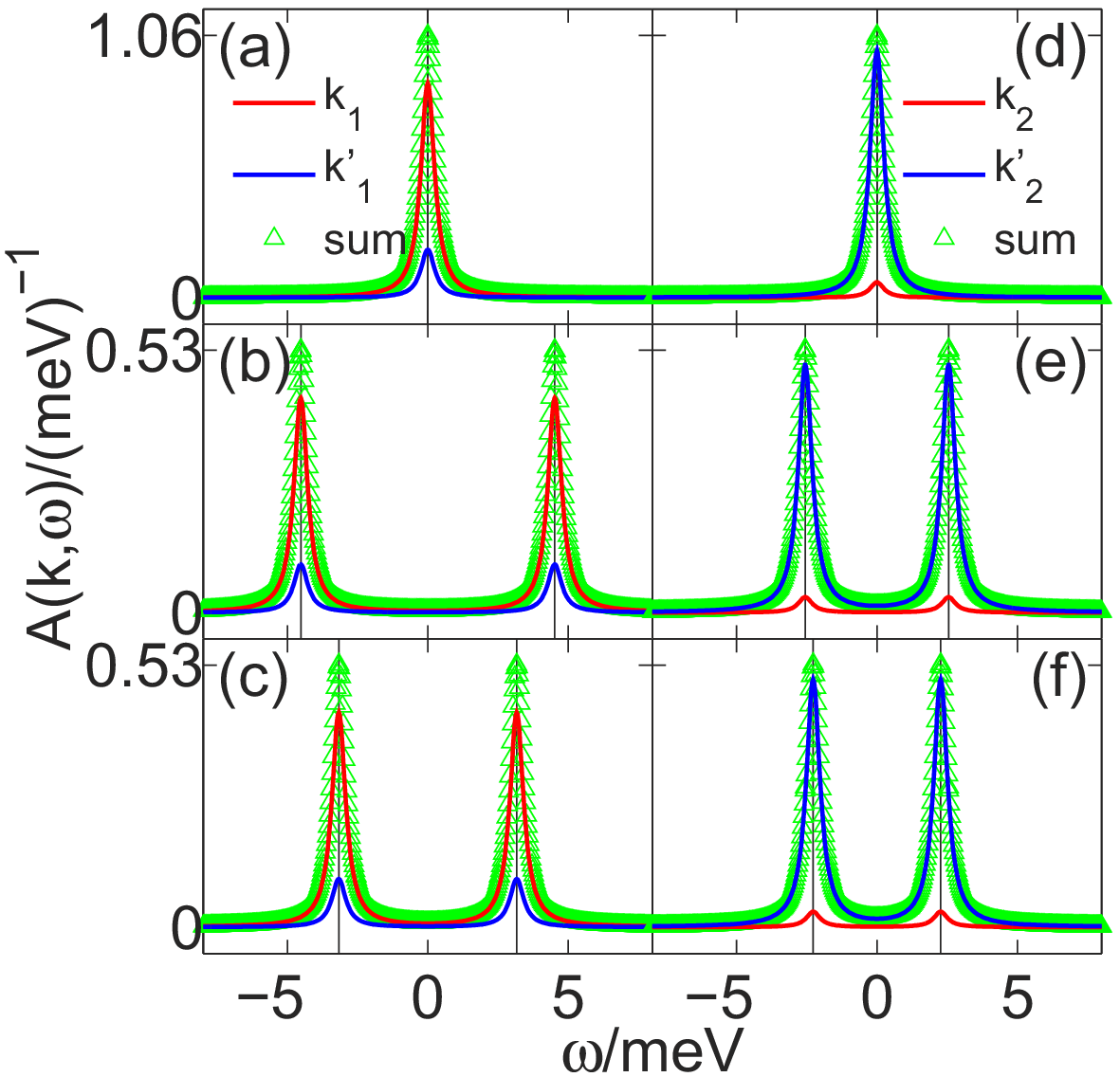}

\caption{Spectral function on the Fermi surface in physical crystal momentum space. Panels (a) and (d) are
for the normal state, (b) and (e) for the superconducting state, and (c) and (f) for the superconducting
state with the $\eta$ contribution removed. Panels (a)--(c) are for momenta $\bk_{1}$ and $\bk'_{1}$ and
(d)--(f) for $\bk_{2}$ and $\bk'_{2}$. The momenta are indicated on the Fermi surface in
Fig.~\ref{fig:orbWt}.}
  \label{fig:LaOFeAs_n5.95_spectraWt}
\end{figure}

\begin{figure}
  \centering
  \includegraphics[keepaspectratio=true,scale=1.0]{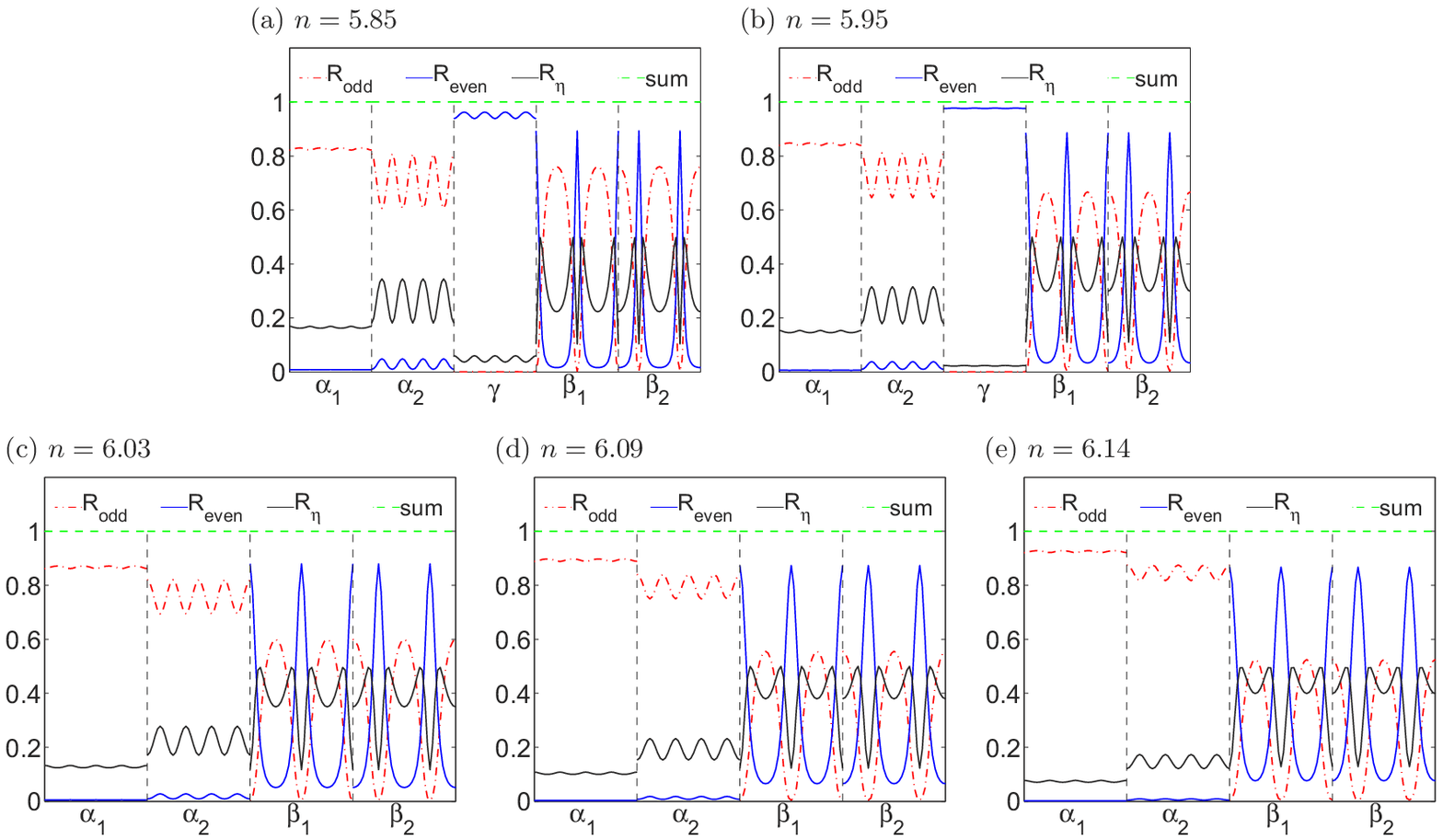}

\caption{(a)--(e): Ratios of $\Delta^N_\text{odd}(\bk)/\tilde{\Delta}_\nu(\bk)$ (red dash-dotted line),
$\Delta^N_\text{even}(\bk+\bQ)/\tilde{\Delta}_\nu(\bk)$ (blue solid line) and
$[\Delta^\eta_\text{odd-even}(\bk)+\Delta^\eta_\text{even-odd}(\bk+\bQ)]/\tilde{\Delta}_\nu(\bk)$ (black
solid line) for five different doping levels. Here $\bk$ is along each Fermi pocket starting from $+k_x$
direction. Top row: hole doping with $\gamma$ pocket. Bottom row: electron doping without $\gamma$ pocket.}
  \label{fig:dopings}
\end{figure}

\begin{figure}
  \centering
  \includegraphics[keepaspectratio=true,scale=1.0]{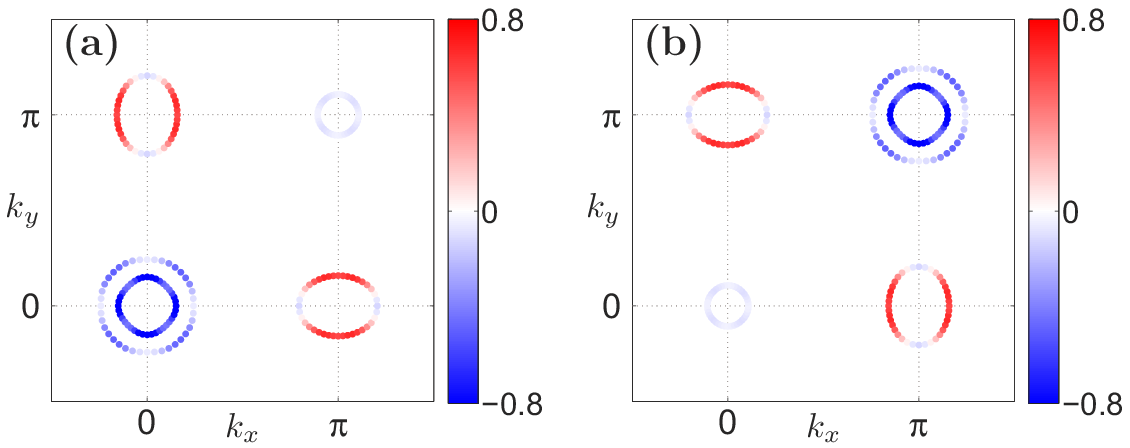}

\caption{$\tilde{\Delta}(\bk)$ calculated at filling $n=5.95$ for $J=0$ on Fermi pockets in physical crystal
momentum space with odd (a) or even (b) orbital character.}
  \label{fig:od_ev_gap}
\end{figure}

\section*{S2. Doping dependence of the normal and $\eta$ contributions to the gap}
As shown in the main text in Eq.~(\ref{eq:Deltasum}), the gap function $\tilde{\Delta}_\nu(\bk)$ calculated
in the five-orbital model in the pseudocrystal momentum representation splits into normal
($\Delta^N_\text{odd}$ and $\Delta^N_\text{even}$) and $\eta$-pairing terms ($\Delta^\eta_\text{odd-even}$
and $\Delta^\eta_\text{even-odd}$) in the physical crystal momentum space according to
\begin{align}
	\tilde{\Delta}_\nu(\bk) = \Delta_\text{odd}^N(\bk)+\Delta_\text{even}^N(\bk+\bQ)
            + \Delta^\eta_\text{odd-even}(\bk) + \Delta^\eta_\text{even-odd}(\bk+\bQ)\,.
\end{align}
Fig.~\ref{fig:dopings} shows the ratios $R_\text{odd} = \Delta^N_\text{odd}(\bk)/\tilde{\Delta}_\nu(\bk)$,
$R_\text{even} = \Delta^N_\text{even}(\bk+\bQ)/\tilde{\Delta}_\nu(\bk)$ and $R_{\eta} =
[\Delta^\eta_\text{odd-even}(\bk) + \Delta^\eta_\text{even-odd}(\bk+\bQ)]/\tilde{\Delta}_\nu(\bk)$ for five
different filling levels $n$. With increasing filling $n$, one sees that the relative contribution from the
$\eta$ pairing decreases on the hole pockets and increases along the $\pm \pi/4$ directions on the electron
pockets. The overall change with doping is found to be rather weak and the $\eta$ pairing remains significant
for all doping levels.

\section*{S3. Superconducting gap with nodes}

The superconducting gap structure we have analyzed in the main text does not have nodes on the Fermi surface.
Here we present the results of a calculation for different parameters (filling $n=5.95$, $U=1.3\ut{eV}$ and
$J=0$), for which we find an $s_\pm$ gap $\tilde{\Delta}({\bf k})$ in pseudocrystal momentum space that has
nodes on the electron pockets. Fig.~\ref{fig:od_ev_gap} shows this gap structure plotted on the ``odd'' (a)
and ``even'' (b) Fermi surface pockets, i.e., as it enters the spectral function. Since the gap is given by
$\tilde{\Delta}(\bk)$ on the ``odd'' Fermi surfaces, but $\tilde{\Delta}(\bk-\bQ)$ on the ``even'' pockets,
the nodes on the electron pockets appear along different directions in the physical crystal momentum space.
This is similar to what was discussed by C.-H.~Lin \textit{et al.}~in Ref.~\onlinecite{Lin2014}.

\end{widetext}


\begin{thebibliography}{00}
\bibitem{footnote}%
To be more precise, $\Delta^N_\text{odd}({\bf k})$ and $\Delta^\eta_\text{odd-even}({\bf k})$ have the same
sign as $\tilde{\Delta}({\bf k})$, and $\Delta^N_\text{even}({\bf k})$ and $\Delta^\eta_\text{even-odd}({\bf
k})$ have the same sign as $\tilde{\Delta}({\bf k}-{\bf Q})$.
\end{thebibliography}
\end{document}